# PAPR Reduction of OFDM Through Pilot Shifting

Md Sakir Hossain
Graduate School of Science and Engineering
Saitama University
Saitama, Japan
Email: sakir@sie.ics.saitama-u.ac.jp

Tetsuya Shimamura
Graduate School of Science and Engineering
Saitama University
Saitama, Japan
Email: shima@sie.ics.saitama-u.ac.jp

*Abstract*—Peak to Average Power Ratio (PAPR) of Orthogonal Frequency Division Multiplexing (OFDM) is a long-standing problem which has been hindering its performance for decades. In this paper, we propose a new PAPR reduction scheme based on shifting pilot locations among the data symbols. Since no side information is sent to the receiver about the pilot locations, a novel pilot detection algorithm is devised here exploiting the pilot power and the relative constant distance property of pilots. The proposed scheme attains around 1.5 dB PAPR reduction. The pilot detection accuracy is shown to be very excellent ranging from 80% to 99% at 0 dB of Signal to Noise Ratio (SNR) in different parameters. This scheme is very spectrally efficient with reduced complexity without degrading BER performance significantly.

*Keywords—OFDM; PAPR; pilot; detection; SLM*

## I. INTRODUCTION

The demand of high data rate requirement is attributed to the massive development of Multimedia communication in last decade. This need of high data rate can be accommodated by using broader spectrum. Unfortunately, the frequency spectrum is a limited resource and becoming scarce as days go on. For this reason, the future of communication systems is largely depending on the efficient utilization of this limited resource. Orthogonal Frequency Division Multiplexing (OFDM) has been considered as the most spectrally efficient physical layer technology. This technology has some other prominent advantages like immunity to Inter Symbol Interference (ISI) and frequency selective fading. This simplifies its receiver design, thereby making OFDM-based devices inexpensive. Because of these merits, OFDM has already been used in applications like Wireless Local Area Network (WLAN), Digital Audio Broadcasting (DAB), Digital Video Broadcasting (DVB), Worldwide Interoperability for Microwave Access (WiMAX), and Japanese Multimedia Mobile Access Communications (MMAC). In addition, OFDM-based multiple access technology, named Orthogonal Frequency Division Multiple Access (OFDMA), is being used in the downlink of fourth-generation (4G) mobile communication technology, named Long Term Evolution (LTE). However, all these advantages do not come alone; these are accompanied by a problem, named Peak-to-Average Power Ratio (PAPR), which hinders its application very badly. The large difference between peak power and average power of the signal compels the High Power Amplifier (HPA) to be operated in a non-linear region which is not desirable in terms of power efficiency and Bit Error Rate (BER) performance. Designing a HPA having a large linear region can eliminate this problem but such a HPA consumes a significant budget of transmitter design thereby making the transmitter expensive.

There are many algorithms already proposed to solve this problem. These include Selected Mapping (SLM) [1], Partial Transmit Sequence (PTS) [2], Companding [3] , Tone Injection (TI) [4], Tone Reservation (TR) [4], and amplitude clipping [5]. A detailed survey of PAPR reduction algorithms can be found in [6]. Most of the PAPR reduction algorithms perform some operations on the data sequence to be transmitted and require sending information about the operation to the receiver as Side Information (SI). For example, SLM involves multiplication of the data sequence by a set of phase sequence and selection of the resulting sequence which provides the lowest PAPR; the phase sequence which provides the lowest PAPR is transmitted as SI. These techniques perform well in reducing PAPR. However, the transmission of SI reduces spectral efficiency, and hence data rate. In addition, since the recovery of the original data sequence largely depends on SI, the need for SI escalates the BER due to the contaminated SI, introduced by the communication channel. Because of these setbacks, PAPR reduction algorithms without the necessity of sending SI is drawing attraction of researchers. Research has been carried out to exploit non-data carrying subcarriers to reduce PAPR. For example, in [7] PAPR is reduced by iteratively exchanging each data subcarrier location with the null subcarrier. This algorithm requires extensive search for finding the optimum data subcarrier. An attempt to reduce search complexity is done in [8]. Nevertheless, such a algorithm remains practically infeasible due to its very poor BER performance at low Signal-to-Noise Ratio (SNR) for constellation size greater than 2. Several researchers exploited pilot tones for the purpose of PAPR reduction. Devlin et el. [9] uses pilot subcarriers and null subcarriers for carrying the phase and amplitudes of clipped data subcarriers, respectively. This approach can effectively reduce out of band radiation compared to the conventional clipping techniques. However, it suffers from almost similar BER performance like clipping. Instead of performing clipping, Hosokawa et el. [10] utilizes the phases of pilot tones. It performs exhaustive searching for optimum pilot subcarrier phases which causes reduced PAPR, which is followed by reducing pilot power exploiting the found phases. Since channel estimation techniques only require pilot power, such an approach does not affect BER performance and requires no SI. However, this approach suffers extremely due to its high computational complexity for optimum pilot phase searching. Changing pilot position as well as phase is used to reduce PAPR in [11]. The pilot position is selected 'pseudo-randomly' among the available subcarriers to obtain the lowest PAPR, thereby degrading channel estimation performance because pilots should be equi-spaced and equi-powered [12]-[13]. This algorithm cannot maintain the equal relative distance between consecutive pilots. The use of multiple levels of pilot power to make each subcarrier equiprobable also affects the channel estimation. No pilot detection algorithm is proposed in [11]. Venkatasubramanian et el. claims to use the pilot detection algorithm provided by Chen [14] at the receiver. However, the Chen's algorithm works only if the distance between consecutive pilots is fixed and known to the receiver. Pilot locations in [11] are chosen randomly just ensuring each subcarrier equiprobable as pilot location. However, making each subcarrier equiprobable does not ensure equidistance between consecutive pilot subcarriers. For this reason the pilot detection at the receiver is not possible using the algorithm in [14]. This makes the PAPR reduction algorithm in [11] useless. The algorithm is not investigated with respect to BER performance as well. Moreover, very high pilot power used to facilitate pilot detection gives rise of average power of the signal considerably which causes an increase in hardware design cost to make a HPA having a large saturation region

In this paper, we propose a new PAPR reduction algorithm which uses pilot location as the only tool to reduce PAPR. Instead of choosing pilot location randomly, we shift pilot symbols among the data symbols so as to maintain equal distance among the pilots. Randomly changing phases or pilot location involves computational complexity for generating random number. The proposed algorithm avoids such complexity by shifting pilot symbols among the data symbols. This shifting does not require any arithmetic operation. Moreover, the equipower property is also maintained. Usually in the pilot tone assisted modulation [12]-[13], the power assigned to the pilot symbols is greater than that of the data symbols. We utilized this property here. However, to contain the average power of the signal, a suitable pilot power which is, of course, greater than that of data symbols is used in this paper. At the receiver side, a pilot detection algorithm is proposed to detect the pilot symbols. This algorithm utilizes the known equal relative distance among the pilots as well as slightly higher pilot power to the detect pilots. This detection algorithm performs well even at low SNR and does not cause significant BER degradation. Although this detection algorithm increases the complexity of the receiver, it paves the way for a spectrally efficient PAPR reduction algorithm.

## II. PROBLEM FORMULATION

OFDM system splits the whole bandwidth into a number of narrow subbands such that the channel gain remains constant over each subband. In addition, each subcarrier is made orthogonal with respect to the rest of the subcarriers. For $N_s$ modulated symbols, the OFDM symbol of length $N_s$ is computed using Inverse Fast Fourier Transform (IFFT) as

$$x_n = \frac{1}{\sqrt{N_s}} \sum_{k=0}^{N_s-1} X_k e^{j2\pi kn/N_s} \quad (1)$$

where $X_k$ and $x_n$ are the frequency and time domain symbols, respectively. The large peak occurs when many sinusoids get added at the same phase and causes a large difference between itself and the average power. PAPR of an OFDM symbol is measured using the following formula:

$$PAPR = \frac{\max(|x_n|^2)}{E\{|x_n|^2\}} \quad (2)$$

where $|x_n|^2$ is the instantaneous power at $n^{th}$ time instant and $E\{.\}$ denotes the expected value over the OFDM symbol period.

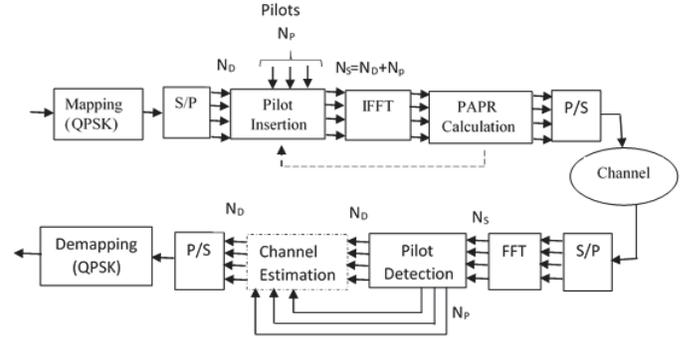

Fig. 1 Block diagram of proposed OFDM system.

However, some of the peaks remain missed when $N_s$ samples are taken in a discrete-time baseband signal [6]. For this reason, the PAPR computed in a discrete-time baseband signal is smaller than that taken from the corresponding continuous-time baseband signal. To overcome this problem, $L(L \geq 4)$ times oversampling is done to measure PAPR more precisely from the discrete-time baseband signal. In that case, $N_s$ in (1) is replaced by $N_s L$.

## III. PROPOSED TECHNIQUE

In this section, we describe our proposed OFDM system. We propose modifications of transmitter as well as receiver. In the transmitter, the positions of pilot symbols are changed iteratively to select a pilot arrangement that provides the lowest PAPR. In the receiver, we propose a pilot detection algorithm for efficient pilot identification. The novelty of the proposed system involves keeping the distance between consecutive pilots unaffected and using the same level of power for all pilots, thereby making pilot detection at the receiver feasible.

### A. Transmitter Design

A block diagram of the proposed system is shown in Fig. 1. Before the serial-to-parallel (S/P) conversion, $N_D$ data symbols are mapped to a particular constellation. The pilot insertion block accepts $N_s = N_D + N_p$ symbols as input in which $N_p$ represents the number of pilot symbols. The function of this block is to insert pilot symbols of equal power in different locations maintaining constant distance between consecutive pilots. Initially, a pilot symbol is kept in front of the data symbol sequence of length $N_D$ and other pilots are set at a constant interval in the following sequence, thereby making a pilot and data sequence of length $N_s$ in the OFDM signal. An example of initial pilot insertion between the data symbols is shown in Fig. 2(i) in case of $N_s = 8$ and $N_p = 2$. PAPR of this arrangement of pilots and data symbols is computed after IFFT operation. This sequence along with the corresponding PAPR are stored. Then, the process goes back to the pilot insertion block (the dashed arrow in Fig. 1 indicates iterations). The same number of pilots are inserted in the pilot insertion block in such a way that the new location of each pilot is just one position right to that of its previous position, that is, the index of each pilot is increased by 1. The resulting arrangement looks like Fig. 2(ii). It is evident from Fig. 2(i) and Fig. 2(ii) that the pilots of the latter ones in Fig. 2(iii) and 2(iv) are shifted versions of each former one. PAPR of the corresponding time-domain signal is computed and stored like before. These steps of inserting pilots into one position right to their previous positions, followed by IFFT and PAPR calculation continue until PAPR is reached below a threshold or $r_o$ becomes equivalent to $R$. The time domain symbol sequence after IFFT, with a particular pilot arrangement, which provides the lowest PAPR is transmitted after the parallel-to-serial (P/S) conversion. In the proposed system, no information about the pilot locations is sent to the receiver and one relies on the receiver that detects pilot locations using the pilot detection algorithm described to be in the following subsection.

In general, the distance among two consecutive pilots' indexes is defined by

$$R = \frac{N_s}{N_p} \quad (3)$$

The locations of the $i$-th pilot, $\beta_i$, is given by

$$\beta_i = K_i R + r_o \quad (4)$$

where $K = [K_1, K_2, K_3, \ldots, K_{N_p}] = [0, 1, 2, \ldots, N_p - 1]$, and $r_o$ is the location of the first pilot among all symbols and satisfies $0 < r_o \leq R$ ($r_o$ is an integer).

*B. Receiver Design*

Due to the change of pilot locations in the transmitted data sequence, the receiver of the proposed OFDM system is different from that of the conventional systems and a measure needs to be taken to identify the pilot symbols. Large pilot power may be used to facilitate pilot identification but such high power causes a rise of average signal power which results in a surge of hardware design cost. The proposed pilot detection algorithm detects the pilots having comparatively lower pilot power with a detection accuracy that does not affect BER significantly.

The algorithm to detect pilot symbols utilizes larger power of the pilot symbols relative to the data symbols and equal distance between the consecutive pilots. The pilot detection is performed on the frequency domain signal obtained by computing FFT of

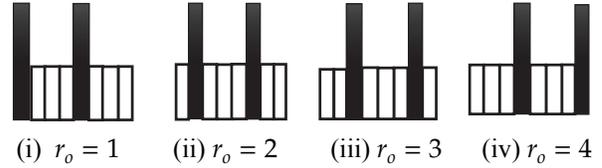

(i) $r_o = 1$ (ii) $r_o = 2$ (iii) $r_o = 3$ (iv) $r_o = 4$
Fig. 2 Shifting of pilot symbols (black bars) among the data symbols (white bars)

the received time-domain signal. A number of symbols' locations having power greater than a threshold are segregated. Modular additions of a set of numbers, which are integer multiples of $R$, to each of these symbol locations are performed to find $N_p$ number of locations those appear in the largest number of times. $N_p$ number of locations found in this way are the pilot location. However, further measures are taken to improve the performance. A set of pilot locations are computed exploiting the equidistance property of pilots from each pilot locations computed in the previous step, and the sequence of symbol locations which results in the largest aggregated power is finally considered as the desired pilot sequence. Detail procedure is described below.

Let $Y$ be a frequency domain representation of the received time domain symbols' set $y$ as

$$Y = FFT(y). \quad (5)$$

Let $A = [a_1, a_2, a_3, \ldots, a_{N_s}]$ be the set of locations of the elements of $Y$ in an ascending order. There is a one-to-one correspondence between the elements of $Y$ and $A$ by

$$a_i = Y_i \quad (6)$$

for $i = 1,2,3,\ldots,N_s$. Another set of length $N_s$, $B = [b_1, b_2, b_3, \ldots, b_{N_s}]$, is used here. Each element of $B$ is defined as

$$b_i = |Y_i| - \gamma\sqrt{P} \quad (7)$$

where $|Y_i|$ is the absolute value of $i$-th symbol, $P$ is the pilot power, and $\gamma$ is a design parameter which determines how many symbols will be considered as the candidates of pilots, having values $0 < \gamma \leq 1$. Furthermore, another set of length $M$, $Q = [q_1, q_2, q_3, \ldots, q_M]$, is used. For each element of $Q$,

$$q_i = a_j \quad (8)$$

if $B_j > 0$ is satisfied for $\forall j \leq N_s$. The theoretical distances of each pilot symbol from the rest of pilots, which are known to the transmitter and receiver, are given by

$$S_i = T_i R \quad (9)$$

where $T = [T_1, T_2, T_3, \ldots, T_{N_p-1}] = [1, 2, 3, \ldots, N_p - 1]$ and $S = [S_1, S_2, S_3, \ldots, S_{N_p-1}]$. A new matrix $D$, whose elements are $D_{ij}, i = 1,2,3,\ldots,M; \ j = 1,2,3,\ldots,N_p - 1$, is formed combining $Q$ with $S$ as

$$D_{ij} = \begin{cases} v_j mod(N_s + 1) + 1 & ; \quad for \ v_j mod(N_s + 1) < v_j \\ v_j & ; \quad otherwise \end{cases} \quad (10)$$

where $v_j = q_i + S_j$.

If a noise free channel is considered, the $Q$ vector will consist of $N_p$ elements which are the locations of the desired pilots.

Then, each row of $D$ will consist of the locations of $N_p - 1$ pilots located at the distance defined by $S$. The row of a true pilot location will consist of the rest of the pilot locations. Since a noise free channel does not exist in practice, $Q$ includes some of the true pilot locations as well as spurious pilot locations. If an element of $Q$ is a true pilot location, the result of addition of $S$ with that element will consist of those elements of $Q$ which are the true pilot elements as well as the true pilot elements which are not included in $Q$. However, if a spurious pilot location contained in $Q$ is added to $S$, a lesser number of elements of $Q$ will be produced. The following vector $C$ consists of the number of the elements of $Q$ produced from (10) by each element of $Q$;

$$C = [c_1, c_2, c_3, \ldots, c_M]$$

where $c_i$ is the number of elements of $Q$ found in $i$-th row of $D$. The desired pilot locations are determined from $C$. The location of each pilot corresponds to the index from $Q$ which provides one of the $N_p$ largest elements of $C$. This results in a set of initially estimated pilot locations' set $U$ whose each element is defined as

$$u_i = q_j \tag{11}$$

if $c_j$ is among the largest $N_p$ elements of $C$, where $i$ is an integer and ranges $0 < i \leq N_p$.

$U$ may consist of one or more erroneous pilot locations because few pilot symbols power may get reduced significantly, hence not included in $Q$. In addition, some false detection of pilot location may also occur. Thus, checking of whether $U$ contains any falsely detected pilot or whether could not detect the true pilot is imperative to improve the performance. To perform this checking, the relative distance property of the pilot symbols is utilized here. The distance of each pilot from other pilots is of integer multiples of $R$. This property is used to find whether $U$ consists of all the pilots. $N_p$ pilot locations are calculated from each element $u_i$ of $U$ using

$$Z_{ij} = \begin{cases} w_j mod(N_s + 1) + 1 & ; \quad for\ w_j mod(N_s + 1) < w_j \\ w_j & ; \quad otherwise \end{cases} \tag{12}$$

where $w_j = (u_i + nR)$, both $i$ and $j$ are integers and range from 1 to $N_p$, and $n = [1, 2, 3, \ldots, N_p]$. Final pilot locations are determined by computing the total power of each of the rows of $Z_{ij}$ as

$$\alpha_i = \sum_{j=1}^{N_p} |Y_{Z_{ij}}|. \tag{13}$$

The $k$-th row of $Z_{ij}$ is selected as the desired pilot locations' sequence if

$$\alpha_k = max(\alpha_i).$$

Use of a fixed $\gamma$ in (7) may result in failing to capture a sequence of candidate pilot locations that cannot satisfies the equidistance property, particularly at low SNR. To overcome this problem, a soft $\gamma$ can be used in the sense that the $\gamma$ value will be decremented by a predefined amount to capture more candidate pilots and all steps after (7) will be repeated for each $\gamma$ until a pilot sequence is found. This soft $\gamma$ undoubtedly.

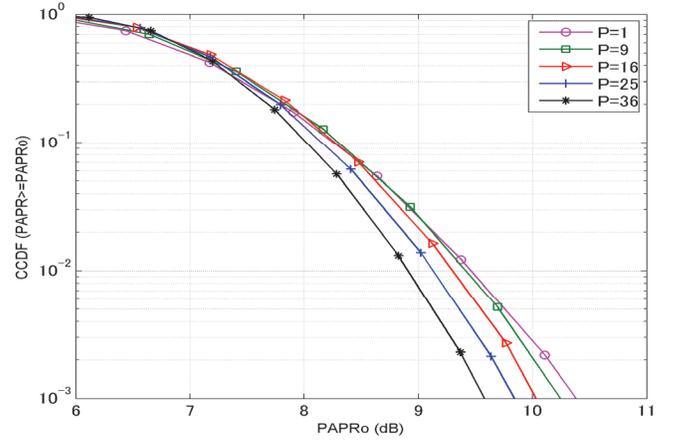

Fig. 3 Effect of pilot power on PAPR

improves the performance at the expense of higher computational complexity

*C. Features of the systems*

The advantages of the proposed OFDM system include 1) it does not require exhaustive search for optimum phase of pilots [10] or pilots' phase and locations like [11] or suitable subcarriers for exchanging with null subcarriers[7]. 2) It is spectrally efficient unlike [16] because no subcarrier is reserved here for sending SI to the receiver or reducing PAPR. 3) Since this is a pre-IFFT process, it does not cause spectral regrowth making out-of-band radiation minimum unlike [9], 4) Since no SI is required to send, low SNR cannot affect BER performance significantly, 5) unlike [11], it does not increase average signal power as it uses much lower pilot power. However, all of these merits are achieved at the cost of increased computational complexity at the receiver. This complexity is introduced by the pilot detection block.

## IV. SIMULATION RESULTS

The performance of the proposed systems is investigated based on the PAPR reduction capability at the transmitter and the pilot detection accuracy at the receiver. Simulations are conducted in baseband systems. Since the PAPR in passband is practically double to that in baseband [15], the simulation result in baseband remains valid in passband as well. Throughout this paper, Quadrature Phase Shift Keying (QPSK) is used as the modulation technique. Guard interval is not inserted at the transmitter because it has no effect on PAPR and BER on a static channel. Each simulation is computed sending $1 \times 10^5$ OFDM symbols. $L = 8$ is used to capture peaks more precisely. Before starting the simulation of the proposed transmitter and receiver, it is necessary to find the proper pilot power to rein in rise of average power significantly. Fig. 3 shows PAPR of a conventional OFDM system with different pilot power at $N_s = 64$ and $N_p = 4$. It is clear that $P \geq 39$ brings about increase of average power so much that the resulting higher average power itself causes PAPR reduction. For this reason, $P = 9$ is considered as the desired pilot power unlike $P = 39$ in [11].

Fig. 4 depicts the PAPR reduction capability of the proposed system using Complementary Cumulative Distribution Function (CCDF), where M.OFDM stands for Modified OFDM, that is, the proposed system. The proposed system reduces PAPR by roughly 1.5 dB and 2.0 dB at the probability of $10^{-3}$ and $10^{-2}$, respectively, in case of $N_p = 4$. Degradation of PAPR reduction capability is observed when $N_p$ is increased to 8. This is due to the fact that the possible number of shifting of pilots goes down by half as $N_p$ increases to 8 from 4. Another observation is that when the number of pilots is increased, the original OFDM itself causes more PAPR due to the higher amplitudes of increased number of subcarriers. Thus, the best performance of the proposed algorithm is achieved with the least number of pilots. However, the increase of total number of pilots does not degrade the system performance as long as the ratio of total number of symbols to the number of pilot symbols is maintained.

The PAPR has been successfully reduced in the transmitting end. Now, the challenge is to detect the pilot symbols at the receiver. In the performance evaluation of the proposed detection algorithm, no oversampling at the transmitter is used because oversampling technique is useful for properly computing PAPR but has no effect on BER unless a HPA is used at the transmitter. To make the system simple, the HPA is omitted from the transmitter. In addition, only a static channel (i.e., Additive White Gaussian Noise (AWGN) Channel) is considered throughout this simulation. Channel estimation is not performed here. Our aim is to just detect the pilot symbols and see the effect of wrong detection on overall BER. BER of the proposed receiver largely depends on the proper detection of pilot symbols. The detection error in percentage at 0 dB of SNR is shown in Fig. 5 which is produced using smoothing spline. $R$ is set to 16. Since the detection algorithm is based on pilot power and their relative distance property, the pilot detection accuracy varies with pilot power; the higher the pilot power, the more precise the pilot detection, which results in lower error rate. This is reflected in the figure. In addition, the error rate depends the parameter $\gamma$ of (7). From Fig. 5, the proper value of $\gamma$ can be between 0.66 to 0.9.

The effect of SNR on the performance of the proposed pilot detection scheme is shown in Table I, which is produced using a soft $\gamma$. It is evident that there is a direct relationship between $R$ and detection accuracy. The percentage of false detection of pilot is less than 5 as long as $R = 16$ but becomes higher when $R$ reduces to 8. This rise of error detection stems from the larger number of candidate pilots making the detection more difficult. However, this false detection rate declines to around 1% with some exception at 3 dB before reaching practically to 0 at 6 dB.

The effect of wrong pilot detection on BER needs to be investigated. Fig. 6 shows the BER of the proposed scheme. $P = 9$ is used in both Original OFDM and proposed scheme. In

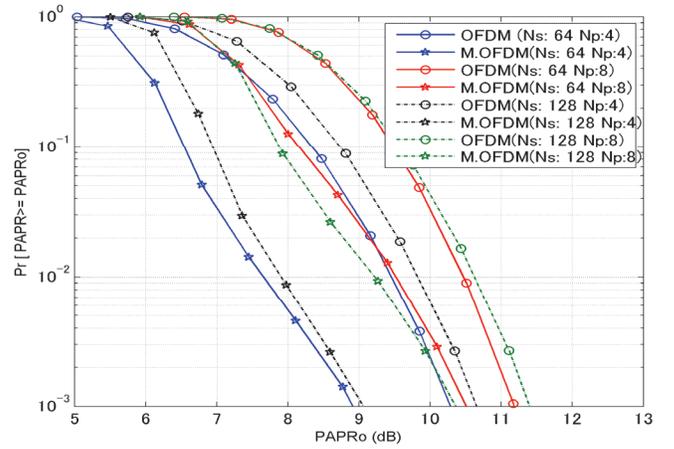

Fig. 4 CCDF for PAPR of the proposed system.

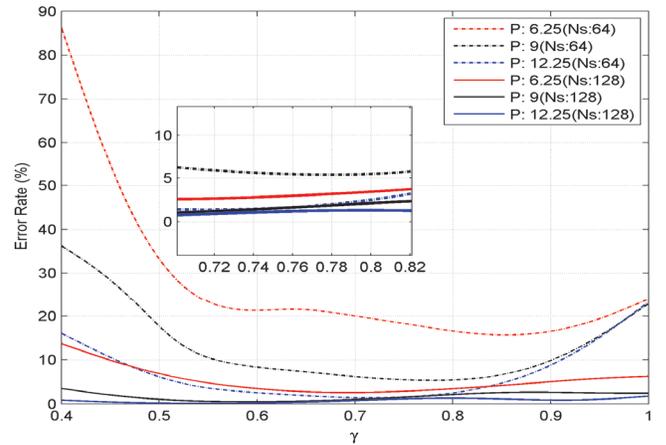

Fig. 5 Percentage of error rates in detecting pilots with respect to various pilot powers.

$P = 9$ is used in both Original OFDM and proposed scheme. In the former case, the pilot location is considered known at the receiver, but unknown (hence, needs to be detected) in the latter one. It is observed that there is a little degradation of the BER performance at low SNR when $R = 8$ and $N_s = 256$ at which pilot detection accuracy is the worst according to Table I. However, there is no noticeable BER degradation at higher $R$. Overall, the proposed scheme does not affect BER in such a way to degrade the system performance.

TABLE I. ERROR DETECTION RATE (%)

| SNR (dB) | | 0 dB | 3 dB | 6 dB | 9 dB |
|---|---|---|---|---|---|
| $N_s = 16$ | $R = 16$ | 4.84 | 0.01 | 0 | 0 |
|  | $R = 8$ | 4.52 | 1.43 | 0.03 | 0 |
| $N_s = 128$ | $R = 16$ | 1.49 | 0.32 | 0 | 0 |
|  | $R = 8$ | 10.89 | 6.42 | 0.23 | 0 |
| $N_s = 256$ | $R = 16$ | 4.27 | 1.84 | 0.01 | 0 |
|  | $R = 8$ | 16.41 | 14.9 | 1.87 | 0 |

V. CONCLUSION

In this paper, a new PAPR reduction algorithm is proposed which solely depends on the locations of pilot symbols. This scheme performs very well at a smaller number of pilots without increasing average signal power. It is a 'blind' scheme in the sense that it does not send any side information to the receiver. The pilot locations are detected at the receiver efficiently . The accuracy rate is more than 90% for the cases of SNR above 3 dB. The detection scheme performs excellently at a small number of pilots and acceptable detection accuracy is observed at very low SNR. It does not degrade BER significantly due to its excellent detection accuracy.

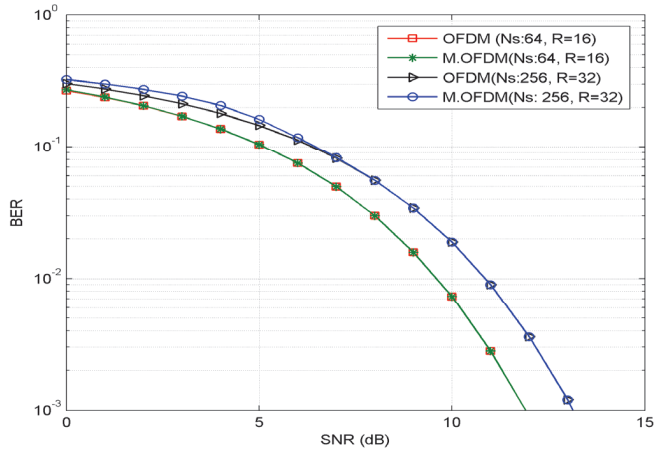

Fig. 6 Comparison in BER of the proposed system with original OFDM